\documentclass[twocolumn,showpacs]{revtex4}

\usepackage{graphicx}
\usepackage{dcolumn}
\usepackage{amsmath}
\usepackage{amssymb}

\newcommand{\dotprod}{{\scriptscriptstyle \stackrel{\bullet}{{}}}}

\begin{document}

\title{Rayleigh-B\'{e}nard Convection in Large-Aspect-Ratio Domains}

\author{M. R. Paul}
 \email{mpaul@caltech.edu}
\author{K-H. Chiam}
\author{M. C. Cross}
\affiliation{Department of Physics, California Institute of Technology 114-36,
Pasadena, California 91125}

\author{P. F. Fischer}
\affiliation{Mathematics and Computer Science Division, Argonne
National Laboratory, Argonne, Illinois 60439}

\date{\today}

\begin{abstract}
The coarsening and wavenumber selection of striped states growing
from random initial conditions are studied in a non-relaxational,
spatially extended, and far-from-equilibrium system by performing
large-scale numerical simulations of Rayleigh-B\'{e}nard convection
in a large-aspect-ratio cylindrical domain with experimentally
realistic boundaries. We find evidence that various measures of the
coarsening dynamics scale in time with different power-law exponents,
indicating that multiple length scales are required in describing the
time dependent pattern evolution. The translational correlation
length scales with time as $t^{0.12}$, the orientational correlation
length scales as $t^{0.54}$, and the density of defects scale as
$t^{-0.45}$. The final pattern evolves toward the wavenumber where
isolated dislocations become motionless, suggesting a possible
wavenumber selection mechanism for large-aspect-ratio convection.

\end{abstract}


\maketitle

\section{Introduction}
Rayleigh-B\'{e}nard convection in large-aspect-ratio domains is a
canonical system in which to study the emergence of order from
initial disorder in a spatially extended system that is driven
far-from-equilibrium~\cite{cross:1993}. A complete understanding of
the transient dynamics of the emerging order and the long-time
selected pattern is still lacking. In this Letter we investigate the
emergence of striped states when a convection layer is quenched into
an ordered state from random initial conditions. Although much has
been learned for systems of stripes approaching an equilibrium state
(relaxational dynamics), much remains unclear for driven systems that
are approaching a steady non-equilibrium state (non-relaxational
dynamics). This is our focus here.

An important physical property of the final selected pattern is the
spatial wavenumber of the convection rolls. For relaxational systems
the long-time asymptotic state is the one that minimizes the free
energy of the system (or a frozen disordered state if the optimal
state is energetically difficult to reach). For non-relaxational
systems however, the long-time asymptotic state is not one minimizing
a free energy, hence raising the issue of wavenumber selection. Many
wavenumber selection mechanisms have been identified for highly
controlled situations, often limiting the type and number of pattern
defects that interact (for example, selection by grain boundaries,
dislocations, or regions of large curvature) or for particular
pattern geometries (such as axisymmetric convection or spatial ramps
in plate separation)~\cite{catton:1988,cross:1982,paul:2002:pre}.
However, an understanding of the wavenumber selected in a
large-aspect-ratio domain initiated from small random thermal
perturbations remains elusive. Therefore, in a non-relaxational
system such as convection, the long-time asymptotic state is unknown
\textit{a priori} and can be one of an infinite number of ordered
states. The effect on the coarsening dynamics is not currently
understood and is discussed further below.

Substantial work has been done on the pattern coarsening in
relaxational systems that occurs as domains of uniform stripes
compete and grow in size. In this case, the dynamics can be
understood in terms of the monotonic decrease of the free energy.
This provides a useful tool to look for important dynamical
interactions and has been exploited for the case of diblock
copolymers~\cite{harrison:2002}. Experiments using diblock copolymers
have been performed in extremely-large-aspect ratios with more than
$10^5$ microdomain repeat spacings, effectively eliminating boundary
effects, and for durations long enough to reach striped states free
of defects. The orientational correlation length, $\xi_o$, was found
to grow in time as $\xi_o \sim t^{1/4}$, and the dominant coarsening
mechanism was determined to be annihilation events involving
disclination quadrupoles.

The Swift-Hohenberg equation (SH), which is relaxational, and the
Generalized Swift-Hohenberg equation (GSH), which can be either
relaxational or non-relaxational depending on the choice of the
nonlinearity, have been studied as model systems for the coarsening
of striped patterns in periodic geometries. For the SH equation a
measure of the translational correlation length, $\xi_T$, was found
to vary as $t^{1/5}$ in the absence of noise and as $t^{1/4}$ in the
presence of noise~\cite{elder:1992}, although recent deterministic
simulations performed close to threshold in the absence of noise give
$t^{1/3}$~\cite{boyer:2001}. A study of the SH equation and a
non-relaxational GSH equation found that the domain size scaled as
$t^{1/5}$ in all cases~\cite{cross:1995}. However, the
non-relaxational results gave an orientational length scale given by
$\xi_o \sim t^{1/2}$, and the stripe patterns were found to evolve
toward a final wavenumber, $q_d$, where isolated dislocations become
stationary.

To date, experiments on the coarsening dynamics of a
far-from-equilibrium, spatially extended non-relaxational system have
been performed only for the electroconvection of a liquid nematic
crystal~\cite{purvis:2001}. Here the pattern is asymmetric, with
convection rolls forming zig and zag rolls at an angle $\pm \theta$
relative to the nematic anisotropy direction. For this system, using
isotropic measures of the domain growth, it has been found that the
domains grow as $t^{1/5}$ and the domain wall length grows as
$t^{1/4}$~\cite{purvis:2001}.

Coarsening experiments on Rayleigh-B\'{e}nard convection in a
large-aspect-ratio container have not been conducted. A considerable
experimental difficulty is in achieving a spatially uniform initial
state composed of random perturbations; slight variations in the
apparatus will influence the initial pattern emerging from the
disorder. Numerical simulations are free of these difficulties,
however, allowing us to investigate the coarsening dynamics of
Rayleigh-B\'{e}nard convection in an experimentally realistic
geometry for the first time.

\section{Discussion}

We study numerically, using a parallel spectral element
code~\cite{fischer:1997}, Rayleigh-B\'{e}nard convection in a
large-aspect-ratio cylindrical domain (see~\cite{paul:2003} for
related applications). The aspect ratio is a measure of the spatial
extent of the system and for a cylindrical geometry is defined as
$\Gamma = r/d$, where $r$ is the radius of the convection cell and
$d$ is the depth. The Boussinesq equations that govern the dynamics
of the velocity $\vec{u}$, temperature $T$, and pressure $p$, fields
are,
\begin{eqnarray*}
  {\sigma}^{-1} \left(
  {\partial}_t + \vec{u} \dotprod \vec{\nabla} \right) \vec{u}
  &=&  -\vec{\nabla} p + RT \hat{z} + \nabla^2 \vec{u}  , \\
  \left( {\partial}_t + \vec{u} \dotprod \vec{\nabla} \right) T
  &=& \nabla^2 T  , \\
  \vec{\nabla} \dotprod \vec{u} &=& 0,
\end{eqnarray*}
where time differentiation is given by $\partial_t$, $\hat{z}$ is
a unit vector in the vertical direction, $\sigma$ is the Prandtl
number, and $R$ is the Rayleigh number. The equations are
nondimensionalized using the layer depth $d$, the vertical
diffusion time for heat ${\tau}_v$, and the constant temperature
difference across the layer $\Delta T$, as the length, time, and
temperature scales, respectively. All bounding surfaces are no
slip, the lower and upper surfaces ($z=0,1$) are held at constant
temperature, and the sidewalls are perfectly insulating.
\begin{figure}[tbh]
\begin{center}
\includegraphics[width=3.0in]{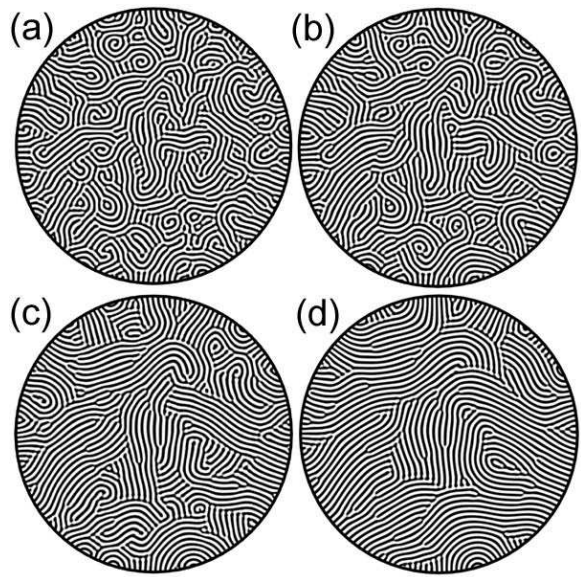}
\end{center}
\caption{The mid-depth temperature field, $T$, at times $t=16,32,64$
and $128$, panels (a)-(d), repsectively. Light regions indicate warm
rising fluid, and dark regions indicate cool descending fluid.
Simulations parameters: $\epsilon=0.27$, $\Gamma = 57$, and $\sigma =
1.4$.} \label{fig_tp}
\end{figure}

We study the pattern evolution from small random thermal
perturbations, $\delta T \sim 0.01$, in a large-aspect-ratio domain,
$\Gamma=57$, containing a fluid with $\sigma=1.4$. We present results
for two simulations at $\epsilon=0.27$ (where $\epsilon = (R -
R_c)/R_c$ is the reduced Rayleigh number and $R_c$ is the critical
Rayleigh number) that differ only in the particular choice of random
initial conditions. Figure~\ref{fig_tp} illustrates the time
evolution of the temperature field at mid-depth for these parameters.
At early times (see Fig.~\ref{fig_tp}(a)) there are present many
small patches of arbitrarily oriented rolls, as well as many defects,
including disclinations, dislocations, grain boundaries, spirals,
wall foci, and targets. As time progresses, the pattern coarsens into
larger domains of stripes with fewer defects mostly dominated by wall
foci, grain boundaries, and isolated dislocations.

\begin{figure}[tbh]
\begin{center}
\includegraphics[width=3.0in]{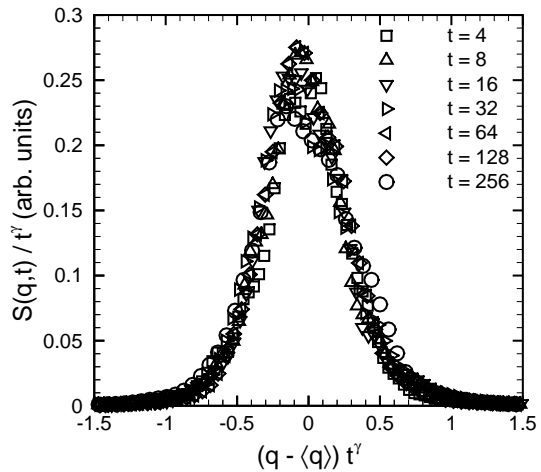}
\end{center}
\caption{The scaling property of the azimuthally averaged structure
factor $S(q,t)$ is illustrated by the data collapse found at various
times when plotting $S(q,t)/t^\gamma$ versus $(q-\left< q \right>
)t^\gamma$ with $\gamma=0.12$.} \label{fig_collapse}
\end{figure}
\begin{figure}[tbh]
\begin{center}
\includegraphics[width=3.0in]{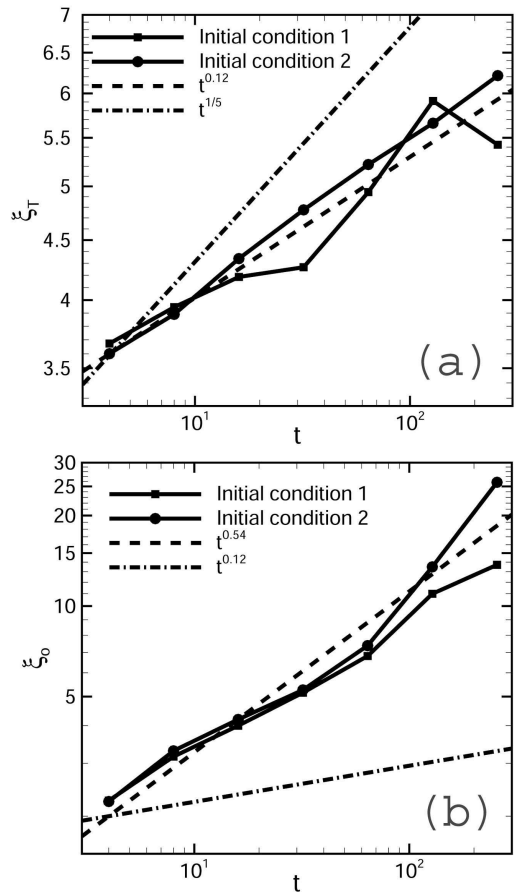}
\end{center}
\caption{Panel~(a): The translational correlation length $\xi_T$ as a
function of time. The dashed line is a power-law fit yielding a
scaling of $t^{0.12}$. The dash-dotted line illustrates a scaling of
$t^{1/5}$ for reference. Panel~(b): The orientational correlation
length $\xi_o$ as a function of time. The dashed line is a power-law
fit yielding a scaling of $t^{0.54}$. For reference the dashed-dotted
line shows the time variation of $\xi_T \sim t^{0.12}$ from panel~(a)
to illustrate the two different length scales.} \label{fig_tocorr}
\end{figure}

A measure of the translational order is the translational correlation
length, $\xi_T$, which is calculated from the time variation of the
second moment of the azimuthally averaged structure factor $S(q,t)$.
The structure factor is the square of the modulus of the spatial
Fourier transform of the temperature perturbation field at mid-depth.
The scaling property of $S(q,t)$ is illustrated in
Fig.~\ref{fig_collapse} by the data collapse at various times by
plotting $S(q,t)/t^\gamma$ versus $(q-\left<q\right>)t^\gamma$ for
$\gamma =0.12$ (where $\left<q\right>$ is the average wavenumber).
The data collapse occurs over a range of $4 \lesssim t \lesssim 256$
indicating a window of time over which the scaling Ansatz is valid.
For large times $t>256$, the scaling breakdown indicates the
influence of lateral boundaries and finite size effects. As shown in
Fig.~\ref{fig_collapse}, the collapse of the $S(q,t)$ curves at early
time $t=4$ (squares) and late time $t=256$ (circles) are beginning to
show some deviation. In the discussion that follows, we consider the
dynamics only in the scaling regime. The translational correlation
length is shown in Fig.~\ref{fig_tocorr}(a). The scaling of $\xi_T
\sim t^{0.12}$ indicates very slow growth when compared with the
predominance of $t^{1/4}$ and $t^{1/5}$ scalings found in a variety
of other systems as already discussed. Similar results are obtained
from measuring the time variation of the inverse half-width at
half-height of $S(q,t)$.

The time dependence of local orientational order is measured from the
time variation of the orientational correlation length, $\xi_o$. This
is determined by calculating the second moment of the azimuthally
averaged Fourier intensity of $\text{Re}[e^{2i\theta}]$, where
$\theta$ is the local angle of the stripes~\cite{egolf:1998}. As
shown in Fig.~\ref{fig_tocorr}(b), $\xi_o \sim t^{0.54}$, which grows
faster than $\xi_T$, as shown by the long dashed line, suggesting the
presence of an additional length scale in the coarsening dynamics.

\begin{figure}[tbh]
\begin{center}
\includegraphics[width=3.0in]{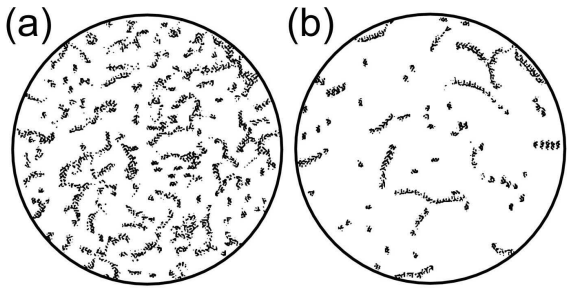}
\end{center}
\caption{Contour plots illustrating the spatial distribution of
defects: panel~(a) $t=16$, panel~(b) $t=128$. Defect regions are
black, and defect-free regions are white. The ratio of the defect
containing area to the total area yields a measure of the defect
density $\rho_d$, which is shown as a function of time in
Fig.~\ref{fig_dcorr}. The patterns corresponding to these defect
distributions are displayed in Fig.~\ref{fig_tp}(a) and~(d).}
\label{fig_defect}
\end{figure}

\begin{figure}[tbh]
\begin{center}
\includegraphics[width=2.5in]{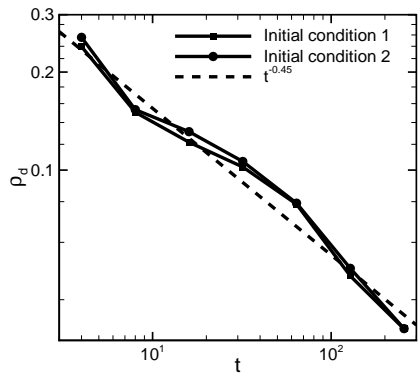}
\end{center}
\caption{The defect density, $\rho_d$, as a function of time. A
power-law fit to the data is shown by the dashed line with $\xi_d
\sim t^{-0.45}$.} \label{fig_dcorr}
\end{figure}

The spatial distribution of defects is quantified in
Fig.~\ref{fig_defect} by highlighting regions of large local
curvature, $\kappa$, where $\kappa = \vec{\nabla} \dotprod \hat{k}$
($\hat{k} = \vec{k}/|k|$ is the local unit
wavevector~\cite{egolf:1998}). There are many defects early in the
time evolution; however, as time progresses, most of the defects are
annihilated, leaving domain walls and isolated dislocations. A defect
density, $\rho_d$, can be defined as the ratio of the total area
covered by defects. The time evolution of $\rho_d$ is shown in
Fig.~\ref{fig_dcorr} and exhibits a scaling of $\rho_d \sim
t^{-0.45}$. For a pattern dominated by isolated defects exhibiting
isotropic growth in all directions, which is approximately valid for
the very early time evolution ($t \lesssim 10$), this suggests a
scaling of the domain size as $\xi_d \sim t^{0.23}$. On the other
hand, for patterns composed of defect lines (or grain boundaries) of
unit width, which is relevant for later times ($t \gtrsim 10$),
$\rho_d$ is the length of the line suggesting a scaling of $\xi_d
\sim t^{0.45}$.

\begin{figure}[h]
\begin{center}
\includegraphics[width=3.0in]{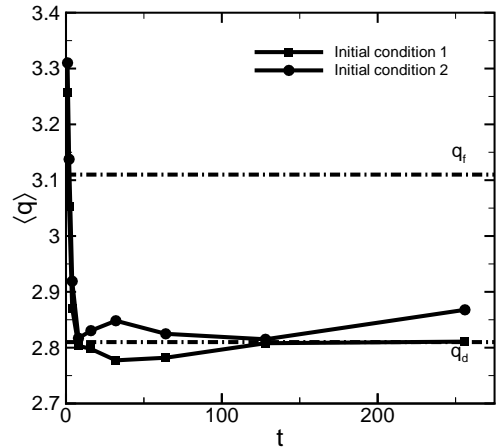}
\end{center}
\caption{Wavenumber variation as a function of time. Wavenumber
selected by zero velocity dislocations $q_d = 2.81$~\cite{plapp:1998}
and wavenumber selected by patches of curved convection rolls $q_f =
3.11$~\cite{buell:1986:axi}.} \label{fig_wn}
\end{figure}

The long-time asymptotic state of a convection pattern free of the
influence of lateral boundaries remains poorly understood. Our
results suggest that the pattern evolves toward the wavenumber where
isolated dislocations become motionless, $q_d$ (see
Fig.~\ref{fig_wn}). The values of $q_d$ have been obtained, for the
fluid parameters of interest here, both experimentally and
numerically by measuring the climb velocity of a dislocation in a
background of either straight parallel rolls or a giant one-armed
spiral and interpolating to find the wavenumber of zero climb
velocity~\cite{plapp:1998}. For reference, the wavenumber selected by
patches of curved rolls or foci, $q_f$, is also
shown~\cite{buell:1986:axi}. The wavenumber $q_f$ is also where
$D_{\perp} \rightarrow 0$ in the absence of mean flow ($D_{\perp}$ is
the diffusion coefficient perpendicular to the wavevector in the
Pomeau-Manville phase equation~\cite{pomeau:1979}). For large times
where the effects of the boundaries are important $256 \lesssim t
\lesssim 500$ we find a slow increase in the wavenumber indicating
that $q_f$ may be selected for at very long times by the prevalence
of curved rolls from large wall foci. In a similar calculation for
slightly more supercritical conditions, $\epsilon=0.46$, the pattern
wavenumber evolves, for times in the scaling regime, to $q=2.58$
where $q_d=2.63$, and $q_f=3.09$, again suggesting a selected
wavenumber of $q_d$. These results indicate that the wavenumber
selected in large-aspect-ratio domains is $q_d$ in agreement with the
predictions made from numerical simulations of the GSH
equations~\cite{cross:1995}.

\section{Conclusion}
We have investigated domain coarsening and wavenumber selection in a
non-relaxational, extended, and far-from-equilibrium system by
performing full numerical simulations of Rayleigh-B\'{e}nard
convection with experimentally realistic boundary conditions. In
non-relaxational systems the long-time asymptotic state is unknown,
thus raising the question of wavenumber selection and the issue of
how this might affect the coarsening dynamics. For
Rayleigh-B\'{e}nard convection we find that multiple length scales
are necessary to describe the pattern evolution in time. The
coarsening dynamics involve the complicated evolution of many types
of defects, making it difficult to identify dominant coarsening
mechanisms responsible for the observed scaling exponents. Further
insight could be gained by studying defect interactions in simpler
prescribed situations. We also find that the pattern selects the
wavenumber where isolated dislocations become stationary, suggesting
that this may be the wavenumber selected from random initial
conditions in the absence of influences from the lateral boundaries.

This research was supported by the U.S.~Department of Energy, Grant
DE-FT02-98ER14892, and the Mathematical, Information, and
Computational Sciences Division subprogram of the Office of Advanced
Scientific Computing Research, Office of Science, U.S.~Department of
Energy, under Contract W-31-109-Eng-38. We would like to gratefully
acknowledge many useful interactions with J.D. Scheel. We also
acknowledge the Caltech Center for Advanced Computing Research and
the North Carolina Supercomputing Center.
\bibstyle{prsty}


\end{document}